\documentclass[aps,prd,twocolumn,
comment, 
superscriptaddress,groupedaddress,
amsmath,amssymb,floatfix,footinbib]
{revtex4}  
\usepackage{float}
\usepackage{graphicx}
\usepackage{bm}
\usepackage[usenames,dvipsnames]{color}
\usepackage{slashed}
\usepackage{hyperref}
\usepackage{slashed}
\usepackage{graphics}
\usepackage{multirow}
\usepackage{tablefootnote}
\usepackage{booktabs}
\usepackage{amsmath}
\usepackage[capitalise]{cleveref}

\begin{document}

\renewcommand{\(}{\left(}
\renewcommand{\)}{\right)}
\renewcommand{\{}{\left\lbrace}
\renewcommand{\}}{\right\rbrace}
\renewcommand{\[}{\left\lbrack}
\renewcommand{\]}{\right\rbrack}
\renewcommand{\Re}[1]{\mathrm{Re}\!\{#1\}}
\renewcommand{\Im}[1]{\mathrm{Im}\!\{#1\}}
\newcommand{\dd}[1][{}]{\mathrm{d}^{#1}\!\!\;}
\newcommand{\del}{\partial}
\newcommand{\nn}{\nonumber}
\newcommand{\ie}{i.e.\,}
\newcommand{\cf}{cf.\,}
\newcommand{\refeq}[1]{Eq.~(\ref{eq:#1})}
\newcommand{\refeqs}[2]{Eqs.~(\ref{eq:#1})-(\ref{eq:#2})}
\newcommand{\reffig}[1]{Fig.~\ref{fig:#1}}
\newcommand{\refsec}[1]{Section \ref{sec:#1}}
\newcommand{\reftab}[1]{Table \ref{tab:#1}}
\newcommand{\order}[1]{\mathcal{O}\({#1}\)}
\newcommand{\fv}[1]{\left(\begin{array}{c}#1\end{array}\right)}%

\def\tcb#1{\textcolor{blue}{#1}}
\def\tcr#1{\textcolor{red}{#1}}
\def\tcg#1{\textcolor{green}{#1}}
\def\tcc#1{\textcolor{cyan}{#1}}
\def\tcv#1{\textcolor{violet}{#1}}
\def\tcm#1{\textcolor{magenta}{#1}}
\def\tcpn#1{\textcolor{pink}{#1}}
\def\tcpr#1{\textcolor{purple}{#1}}
\definecolor{schrift}{RGB}{120,0,0}

\def \mpl{{m_{pl}}}
\def \azeL{{H_0^L}}
\def \azeR{{H_0^R}}
\def \apaL{{H_\parallel^L}}
\def \apaR{{H_\parallel^R}}
\def \apeL{{H_\perp^L}}
\def \apeR{{H_\perp^R}}

\newcommand{\alphas}{\alpha_\mathrm{s}}
\newcommand{\alphae}{\alpha_\mathrm{e}}
\newcommand{\gfermi}{G_\mathrm{F}}
\newcommand{\GeV}{\,\mathrm{GeV}}
\newcommand{\MeV}{\,\mathrm{MeV}}
\newcommand{\amp}[1]{\mathcal{A}\left({#1}\right)}
\newcommand{\wilson}[2][{}]{\mathcal{C}_{#2}^{\mathrm{#1}}}
\newcommand{\bra}[1]{\left\langle{#1}\right\vert}
\newcommand{\ket}[1]{\left\vert{#1}\right\rangle}

\def\be{\begin{equation}}
\def\ee{\end{equation}}
\def\bea{\begin{eqnarray}}
\def\eea{\end{eqnarray}}
\def\bm{\begin{matrix}}
\def\em{\end{matrix}}
\def\bpm{\begin{pmatrix}}
    \def\epm{\end{pmatrix}}

{\newcommand{\lsim}{\mbox{\raisebox{-.6ex}{~$\stackrel{<}{\sim}$~}}}
{\newcommand{\gsim}{\mbox{\raisebox{-.6ex}{~$\stackrel{>}{\sim}$~}}}
\def\mpl{m_{\rm {Pl}}}
\def\gev{{\rm \,Ge\kern-0.125em V}}
\def\tev{{\rm \,Te\kern-0.125em V}}
\def\mev{{\rm \,Me\kern-0.125em V}}
\def\ev{\,{\rm eV}}

\def\Mpl{M_{\rm Pl}}

\title{\boldmath  \color{schrift}{Preheating and gravitational waves in large-field hilltop inflation}}
\author{Diganta Das}
\author{Shreyas Revankar}
\affiliation{Center for Computational Natural Sciences and Bioinformatics,
International Institute of Information Technology, Hyderabad 500 032, India}

\begin{abstract}
The combined Planck, BICEP/Keck Array and BAO measurements of the scalar spectral index and the tensor-to-scalar ratio from the cosmic microwave background observations severely constrain or completely rule out several models of inflationary potentials. On the other hand, the data seems to favor concave potentials over convex ones. In this paper, we study preheating and gravitational waves after inflation in a large-field, regularized  {\it hilltop} potential where inflation takes place in the concave plateau. The inflaton, $\phi$, is coupled to a subdominant scalar field, $\chi$, through a quartic coupling.  After inflation ends, $\phi$ oscillates about the potential minimum and becomes inhomogeneous. The growth of the fluctuation modes, $\delta\phi_k$ and $\delta\chi_k$, in a homogeneous, oscillating background is analyzed in linear perturbation theory, revealing that small modes likely experience broad self-resonance or external parametric resonance. To determine if the resonances are sufficiently strong to cause unstable growth of the modes we perform a lattice simulation. The lattice simulations demonstrate that, although the initial inhomogeneities generate a stochastic gravitational wave background that remains below the present observational limit, the fluctuations do not grow exponentially, and the occupation numbers of $\delta\phi_k$ and $\delta\chi_k$ remain close to zero. 
	
\end{abstract}

\maketitle
\section{Introduction}
The concept of cosmic inflation, an exponential expansion phase of the universe, was partly motivated by the universe's lack of supermassive magnetic monopoles which are predicted by Grand Unified Theories breaking to the $SU(3)\times SU(2)\times U(1)$ structure of the Standard Model. It was soon realized that an inflationary phase explains the uniformity of the cosmic microwave background's temperature across the sky, as well as the puzzle as of why the universe's density has remained so close to its critical density despite fourteen billion years of evolution \cite{Guth:1980zm, Starobinsky:1980te, Linde:1981mu, Albrecht:1982wi, Kazanas:1980tx, Sato:1980ac}. However, inflation's resounding success comes from its prediction of an almost scale-invariant power spectrum (equal power in all scales) of scalar density perturbations sourced by quantum fluctuations \cite{Bardeen:1980kt, Mukhanov:1990me,Kiefer:2008ku, Sudarsky:2009za, Martin:2012ua,Mukhanov:1981xt, Mukhanov:1982nu, Hawking:1982cz, Guth:1982ec, Starobinsky:1982ee,Bardeen:1983qw, Mukhanov:1985rz}. The prediction has been found to be consistent with the observations of the cosmic microwave (CMB) background and the large-scale structure formation. Inflation also generates an almost scale-invariant spectrum of tensor perturbations, sourced by fluctuation in the background metric. The tensor perturbations manifest themselves as stochastic gravitational waves, and their observation is key to discriminating between the inflationary framework and alternative theories. However, present and near-future laser-interferometer detectors operating mostly in the $\approx$10KHz frequency band are insensitive to the gravitational waves from the inflation era, as these waves are typically in the giga-Hertz frequency band. For projected sensitivities of the future GW detectors, see \cite{Ringwald:2020ist,Ringwald:2022xif}. The primordial tensor fluctuations are also responsible for the temperature anisotropies, and the parity-odd (or $B$-mode) polarization fluctuations of the cosmic microwave background \cite{Seljak:1996ti, Kamionkowski:1996zd, Seljak:1996gy}. The $B$-mode polarization fluctuations are the smoking-gun signals of inflation but have so far evaded the existing CMB observatories. 

To drive the exponential expansion of space, inflation requires a field, often modeled as a spatially homogeneous real scalar field $\phi$, called the {\it inflaton}, with a sufficiently flat potential $V(\phi)$. Prior to the onset of expansion, the scalar field's energy density dominated the primordial universe. Rapid expansion occurs when the field slowly rolls down from the flat plateau towards the minimum of the potential. Slow rolling of the field is ensured because, even if the field starts with a large acceleration, the universe's expansion quickly damps it. This is the standard framework for realizing inflation, known as {\it slow-roll}.

In this process, any pre-inflationary energy density is extremely diluted at the end of inflation, leaving the universe cold and empty. Therefore, an important requirement of any inflation framework is that it must leave the universe in a radiation-dominated stage so that Big Bang nucleosynthesis, a process that is well understood and well tested, can begin. The mechanism through which the universe transitions into a radiation-dominated phase is referred to as {\it reheating}. It is achieved through the dissipation of inflaton's energy into the Standard Model fields, which are often modeled as a subdominant scalar field, referred to in this paper as $\chi$. Eventually, the Standard Model fields reach a thermal equilibrium before the onset of the nucleosynthesis processes. 

The traditional approach of studying reheating has been a perturbative framework of energy transfer where the inflaton decays to the subdominant matter field $\chi$ at a rate $\Gamma$ \cite{Abbott:1982hn, Dolgov:1982th, Albrecht:1982mp}. As was soon realized, the perturbative method may not be the most efficient mechanism of energy transfer, since it overlooks an important aspect of the inflaton field. At the end of inflation, the inflaton becomes inhomogeneous and starts coherent oscillations around its potential minimum. These oscillations can drive {\it self-resonance} on itself as well as external {\it parametric resonance} in fields coupled to it leading to rapid particle production \cite{Kofman:1994rk, Kofman:1997yn}. Due to collective behavior, the parametric resonances lead to explosive particle production, which is highly efficient to reheat the universe.  This non-perturbative \cite{Traschen:1990sw, Dolgov:1989us, Kofman:1994rk, Shtanov:1994ce, Kofman:1997yn} stage of energy transfer is referred to as {\it preheating}. The preheating is followed by perturbative reheating and the thermalization of light degrees of freedom.

Over the past decades, a host of inflationary potential models has been explored in the context of preheating and associated phenomena. Following the release of Planck Collaboration's measurements of the anisotropies in the cosmic microwave background (CMB) in 2018 \cite{Planck:2018jri, Planck:2018vyg}, and subsequently after the publication of combined Planck, WMAP, and BICEP/Keck Observations data \cite{BICEP:2021xfz}, a large number of inflationary models has been disfavored \cite{Martin:2013tda, Martin:2013nzq}.  Among the models favored by the CMB data are potentials with a concave shape. In one such model, originally proposed in \cite{Boubekeur:2005zm}, inflation occurs at the maximum or ``hill'' of the potential, which gives the model its name. The hilltop model has been explored in different contexts in \cite{Kallosh:2019jnl, Lin:2019fdk, Kohri:2007gq, Dimopoulos:2020kol, German:2020rpn, Hoffmann:2021vty, Stein:2022cpk, Dutta:2008qn, Wolf:2023uno, Lillepalu:2022knx}. Though the original hilltop model results in successful inflation \cite{Boubekeur:2005zm}, it has one major drawback. The potential is unbounded from below. In such a case the model leads to a universe that ceases to expand and eventually collapses \cite{Kallosh:2019jnl,Felder:2002jk}. The solution is to ``correct" the potential to form a stable minimum. While taking ad-hoc even powers of  unbounded potentials cures the problem \cite{Kallosh:2019jnl, Hoffmann:2021vty, Lillepalu:2022knx}, a systematic method motivated from an effective field theory point of view is to ``regularize' the potential by adding polynomial terms \cite{Hoffmann:2022kod, Wolf:2024lbf}. Studies with unregularized small-field hilltop potential have been carried out in Refs. \cite{Antusch:2015nla, Desroche:2005yt,Brax:2010ai}.

In the present paper, we consider a regularized quadratic hilltop model coupled to a light degree of freedom $\chi$. In Section \ref{sec:background} we describe the model and briefly review the homogeneous background universe. In Section \ref{sec:slowroll}, after summarizing the slow-roll framework, we constrain the model parameters using the latest data from CMB observations. In Section \ref{sec:perturbation}, we study the growth of the perturbations using a linear perturbation theory and analyze the stability and instability regions using Floquet analysis. In Section \ref{sec:lattice}, the dynamics of the fields are further analyzed through a lattice simulation. The work is summarized in section \ref{sec:summary}.

\section{Background Universe \label{sec:background}}
The homogeneous and isotropic background universe is described by the Friedmann-Robertson-Walker (FRW) metric 
\be\label{eq:gmunu}
ds^2  = g_{\mu\nu}dx^\mu dx^\nu = -dt^2 + a(t)^2 \delta_{ij} dx^i dx^j\, ,
\ee
where $i,j = 1,2,3$ and $a(t)$ is the scale factor. In the background universe there exists the inflaton, a spatially homogeneous real scalar field $\phi$ with a potential $V(\phi)$. During inflation, the universe is dominated by the stress-energy tensor of $\phi$
\be\label{eq:tmunu}
T^\mu_{~~\nu} = {\rm diag}(-\rho, p, p, p)\, ,
\ee
where the energy density and the pressure are 
\be
\rho = \frac{1}{2} \dot{\phi}^2 + V(\phi)\, ,\quad p = \frac{1}{2} \dot{\phi}^2 - V(\phi)\, .
\ee
To aid post-inflationary heating through energy dissipation, the inflaton must couple to Standard Model fields. We model the Standard Model matter sector using a real scalar field $\chi$, which couples to the inflaton through a $\xi^2\phi^2\chi^2/2$ term, where $\xi$ is the coupling constant. The inflaton is also minimally coupled to gravity. So, the action of the universe is
\bea\label{eq:action}
S[g_{\mu\nu}, \phi,\chi] &=& \int d^4 x \sqrt{-g} \bigg( \frac{\mpl^2}{2} R -\frac{1}{2} \partial^\mu\phi \partial_\mu \phi \, \nn \\
&-& \frac{1}{2} \partial^\mu\chi \partial_\mu \chi - V(\phi) - \frac{1}{2}\xi^2 \phi^2 \chi^2 \bigg)\, ,\quad
\eea
where $R$ is the curvature scalar, $d^4x \sqrt{-g}$ is the invariant volume element where $g$ is the determinant of the metric, and $\mpl = 1/\sqrt{8\pi G_N}$ is the reduced Planck mass.

During inflation, the $\chi$ field is subdominant and can be neglected in the action. Varying the action $S[g_{\mu\nu}, \phi,\chi]$ with respect to $\phi$ yields a Klein-Gordon equation that determines the evolution of $\phi$ in the homogeneous expanding background
\be\label{eq:phi-evolution-background}
\ddot{\phi} + 3H \dot{\phi} + \frac{\partial V(\phi)}{\partial \phi} = 0\, .
\ee
Here, $H = \dot{a}/a$ is the Hubble constant and the over-dots indicate derivatives with respect to cosmic time. Similarly, varying the action with respect to the metric $g_{\mu\nu}$ yields Einstein's field equations, from which the Friedmann equations governing evolution of the scale factor $a(t)$ are derived, given the choices \eqref{eq:gmunu} and \eqref{eq:tmunu}  
\bea\label{eq:a-evolution-background}
\begin{split}
H^2 &= \frac{1}{3\mpl^2} \bigg( \frac{1}{2}\ddot{\phi}^2 + V(\phi) \bigg)\, ,\\
\dot{H} &= -\frac{1}{2} \frac{\dot{\phi}^2}{\mpl^2}\, . 
\end{split}
\eea

While the equations \eqref{eq:phi-evolution-background} and \eqref{eq:a-evolution-background} describe the background, the inflationary dynamics is determined by the scalar potential $V(\phi)$. In this paper, we consider regularized hilltop potential of the form \cite{Hoffmann:2022kod, Wolf:2024lbf}
\be\label{eq:reg-hilltop}
V(\phi) = V_0 + \sum_{n=2}^{n=q} \frac{a_n}{n!} \bigg( \frac{\phi}{\phi_0} \bigg)^n\, ,
\ee
where $a_n$'s are the expansion coefficients, $V_0$ has mass dimension of 4, and $\phi_0$ has mass dimension 1. Although the order of truncation $q$ can take any value, given the current and the future sensitivity forecast of the CMB experiments, $q=6$ suffices \cite{Wolf:2024lbf}. To ensure that the leading-order term is describes a quadratic model, the coefficient $a_2$ is restricted to $a_2<0$, while $a_6$ is restricted to be positive to stabilize the potential. In Figure \eqref{fig:potential}, the functional form of the potentials is shown for a set of benchmark values of the free parameters to be discussed in the next section.  Inflation occurs in the region of flat plateau, where the field slowly rolls from $\phi \approx 0$  towards the potential minimum. 
\begin{figure}[h!]
\center
\includegraphics[width=0.44\textwidth]{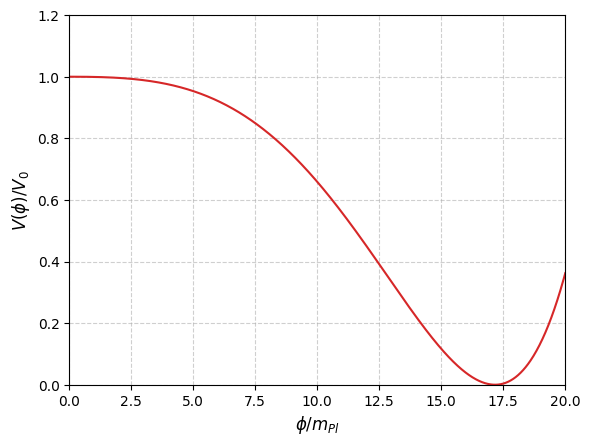}
\caption{The shape of the potential for values given in table. The model parameters are listed in Table \ref{tab:benchmark}.}
\label{fig:potential}
\end{figure}

\section{Slow-roll \label{sec:slowroll}}
During the slow-roll phase of inflation, the kinetic and acceleration terms of the scalar field are vanishingly small ($\dot{\phi}^2\simeq 0$, $\ddot{\phi}\simeq 0$) compared to the potential term. Hence, the equations \eqref{eq:phi-evolution-background} and \eqref{eq:a-evolution-background}  simplify to
\bea\label{eq:slow-roll-dynamics1}
3H\dot{\phi} &\approx& - \frac{\partial V(\phi)}{\partial \phi}\, ,\\
\label{eq:slow-roll-dynamics2}
H^2 & \approx & \frac{V(\phi)}{3m_{\rm Pl}^2}\, .
\eea
The slow-roll scenario is parametrized in terms of the {\it slow roll parameters} $\epsilon_V$ and $\eta_V$, which are defined as
\bea
\epsilon_V &=& \frac{\mpl^2}{2}\bigg( \frac{\partial V/\partial \phi}{V} \bigg)^2 \, ,\\
\eta_V &=& \mpl^2\frac{\partial^2 V/\partial \phi^2}{V} \, .
\eea
The slow-roll phase lasts as long as the slow-roll parameters are small, $\epsilon_V \ll 1$, $|\eta_V| \ll 1$. The end of the slow-roll phase is determined by the field value at which $|\eta_V| \approx 1$, whereas inflation ends at $\phi=\phi_{\rm end}$ when $\epsilon_V(\phi_{\rm end}) =1$. 

The number of e-folds of expansion between the time when observable scales exit the horizon and the time when inflation ends is
\be\label{eq:Nstar}
N_\ast = \frac{1}{\mpl^2} \int_{\phi_{\rm end}}^{\phi_\ast} d\phi \frac{V}{\partial V/\partial \phi}\, ,
\ee
where $\phi_\ast$ is the value of the field when the CMB pivot scale $k_\ast=0.05$Mpc$^{-1}$ exits the horizon. Generally, $N_\ast =50-60$ for a successful inflation. 

The theory so far neglects the role of tiny quantum fluctuations $\delta \phi$ in the inflaton field. These fluctuations produce density perturbations in the background metric. The perturbations are stretched by the inflationary dynamics over cosmological scales. The power spectrum of scalar density perturbations is 
\bea
\mathcal{P}_{\delta\phi} = \langle |\delta \phi(k) |^2 \rangle \propto A_s(k)^{n_s-1}\, ,
A_s = \frac{1}{8\pi^2} \frac{1}{\epsilon_V} \frac{H^2}{m_{\rm Pl}}\, ,
\eea
where $A_s$ is the amplitude, $k$ is the wave number of a mode, and $n_s$ is the scalar spectral index characterizing the scale-dependence of the spectrum. It is defined as
\be
n_s(k) - 1 = \frac{d\ln \mathcal{P}_{\delta\phi}}{d\ln k}\, .
\ee
The proximity of $n_s$ to to 1 in the CMB data (see equation \eqref{eq:CMB-data}) indicates a near scale-invariant scalar power spectrum.

Gravitational waves can be sourced classically by perturbations arising during preheating due to anisotropic stress from inhomogeneities in the matter fields. Independent of the matter fields, the gravitational waves can also be sourced quantum mechanically by the vacuum fluctuations of the tensor metric modes during inflation.
Tensor perturbations also have an almost scale-invariant power spectrum with amplitude
\be
A_t = \frac{2}{\pi^2} \frac{H^2}{\mpl^2}\, .
\ee
From the two amplitudes one can construct an observable called {\it scalar-to-tensor ratio}
\be
r = \frac{A_s}{A_t}\, .
\ee

The three observables $r$ and $n_s$, and $A_s$, can be expressed in terms of the slow-roll parameters at horizon exit as
\bea
n_s &=& 1 + 2\eta_V(\phi_\ast) - 6 \epsilon_V(\phi_\ast) \, ,\\
r &=& 16 \epsilon_V(\phi_\ast) \, ,\\
A_s &=& \frac{1}{24\pi^2 m_{\rm pl}^4} \frac{V(\phi_\ast)}{\epsilon_V(\phi_\ast)}\, .
\eea
The Planck, WMAP, and BICEP/Keck observations data \cite{BICEP:2021xfz} provide the following results on the observables
\begin{align}\label{eq:CMB-data}
\begin{split}
n_s &= 0.96389\pm 0.0043795\, ,\\
r &= 0.030031\pm 0.019744\, ,\\
\ln (10^{10}A_s) &= 3.044 \pm 0.014\, .
\end{split}
\end{align}

We use the Planck dataset to constrain on the model parameters $V_0$, and $\phi_0$, and $a_n$ as shown in figure \ref{fig:planck-ns-r} where each orange dot represents a set of parameters of our model. From the set of constrained parameters, we choose benchmark points shown in table \ref{tab:benchmark} for the rest of our analysis.
\begin{figure}[h!]
\center
\includegraphics[width=0.44\textwidth]{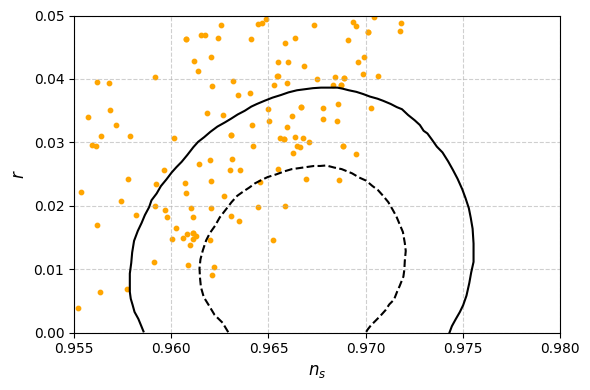}
\caption{Each dot represents a set of model parameter values. The black dotted line (solid line) indicates the 1$\sigma$ (2$\sigma$) allowed region from the combination of Planck, BICEP/Keck, and BAO data \cite{BICEP:2021xfz}. }
\label{fig:planck-ns-r}
\end{figure}
\begin{table}[h!]
\begin{center}
\renewcommand{\arraystretch}{1.} 
\begin{tabular}{|c |c |c |c |c |c |c |}
\toprule\hline\hline 
\midrule\hline 
$V_0$ & $\phi_0$  & $a_2$ & $a_3$ & $a_4$ & $a_5$ & $a_6$ \\
\hline
$66.95$ & 3.76 & $-0.955$ & $-4.41$ & $-5.261$ & $0.231$ & $5.56$ \\
\toprule\hline 
\end{tabular}
\caption{The benchmark values for the model parameters used for the rest of the analysis. The values of $V_0,a_2,a_3,a_4,a_5,a_6$ are reported in terms of $10^{-11}\mpl^4$ and $\phi_0$ is reported in terms of $\mpl$ } 
\label{tab:benchmark}
\end{center}
\end{table}

\section{Linear Preheating \label{sec:perturbation}}
Before delving into the discussion of preheating phase, we present the behavior of the slow-roll parameters $\eta_V$ and $\epsilon_V$ in figure \eqref{fig:epsilon_eta}. As the figure shows, both $\epsilon_V$ and $|\eta_V|$ reach $\sim 1$ almost simultaneously, meaning that the slow-roll phase and inflation ends at the same time.  This contrasts with unregularized small-field hilltop models \cite{Brax:2010ai, Antusch:2015nla, Antusch:2015vna}, where the slow-roll phase ends (when $|\eta_V|\sim 1$), but inflation continues (since $1\ll \epsilon_V$). Another difference from small-field inflation is the absence of {\it tachyonic preheating}. In small-field inflation, it occurs during the period between the end of slow-roll phase and the end of inflation. During this period, the $\partial^2 V/\partial\phi^2<0$, causing sub-Hubble perturbations (long wavelength perturbations) with $k/a<\sqrt{-\partial^2 V/\partial\phi^2}$ experience exponential growth. As shown in figure \eqref{fig:epsilon_eta}, tachyonic preheating is absent in large-field inflation since by the time preheating starts, $\partial^2 V/\partial\phi^2>0$. Although the figure \eqref{fig:epsilon_eta} is shown for a specific benchmark point shown in table \ref{tab:benchmark}, we have verified that the result is true for other benchmark points that are consistent with the Planck data.
\begin{figure}[h!]
\center
\includegraphics[width=0.4\textwidth]{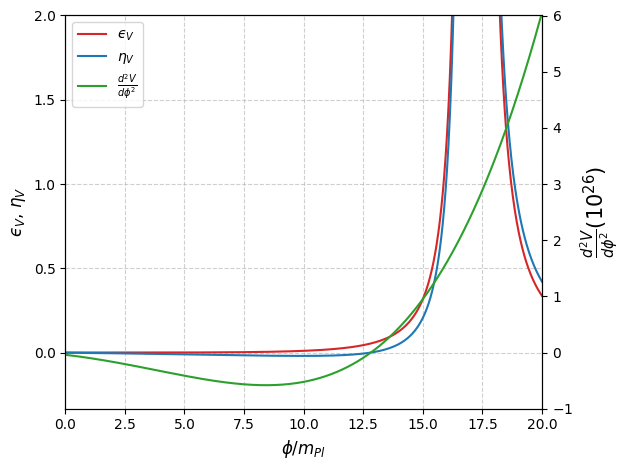}
\caption{The slow-roll parameters $\epsilon_V$, $\eta_V$ and $V^{\prime\prime}(\phi)= \partial^2 V/\partial\phi^2$ as a function of the inflaton field.}
\label{fig:epsilon_eta}
\end{figure}

According to the perturbative description of reheating the universe \cite{Abbott:1982hn,Dolgov:1982th, Albrecht:1982mp}, the individual inflaton particles decay to the auxiliary field $\chi$ at a rate $\Gamma$. The $\phi\to \chi$ production introduces an additional damping term $\Gamma \dot{\phi}$ in the evolution equation \eqref{eq:phi-evolution-background} 
\be\label{eq:phi-evolution-background-withdamping}
\ddot{\phi} + 3H \dot{\phi} + \Gamma \dot{\phi} = - \frac{\partial V(\phi)}{\partial \phi}\, .
\ee
For a small coupling $\xi$ between $\phi$ and $\chi$, which is ensures the validity of the perturbative treatment of calculating $\Gamma$, the damping term $\Gamma\dot{\phi}$ is negligible compared to $3H\dot{\phi}$. Therefore, the energy loss of inflaton due to Hubble expansion is dominates over its energy loss to $\chi$ particles. At later time, as the Hubble constant decreases, $\Gamma$ becomes comparable to $H$, causing the inflaton to decay exponentially reheating to end.

Besides requiring a small coupling, the perturbative description is not the most efficient mechanism of energy transfer since it the inflaton's coherent oscillations near the potential minimum is treated perturbatively.  The oscillation of the homogeneous background amplifies the inflaton's perturbations, leading to self resonance. It also triggers explosive production of $\chi$ through external parametric resonance. Parametric resonance is an efficient mechanism of energy transfer and is described by non-perturbative method \cite{Kofman:1997yn, Shtanov:1994ce, Kofman:1994rk, Traschen:1990sw}.

The equations of motion for the fields following from the action \eqref{eq:action} are 
\bea\label{eq:EoM-phi-chi1}
\ddot{\phi} + 3H \dot{\phi} -\frac{{\boldsymbol \nabla}^2}{a^2} \phi + \frac{\partial V}{\partial \phi} + g^2 \phi \chi^2  &=& 0\, ,\\
\label{eq:EoM-phi-chi2}
\ddot{\chi} + 3H\dot{\chi} -\frac{{\boldsymbol \nabla}^2}{a^2}\chi + g^2 \phi^2 \chi  &=& 0\, .
\eea
At the early stage of preheating, inflaton field is almost homogenous so that the small fluctuations $\delta \phi$ and their growth can be treated in linear perturbation theory. To this end, both the spacetime metric and the scalar fields are expanded to first order around their background values 
\bea\label{eq:perturbation}
\begin{split}
g_{\mu\nu}(t,{\bf x}) & \to & {\bar{g}}_{\mu\nu}(t) + \delta g_{\mu\nu}(t,{\bf x})\, ,\\
\phi(t,{\bf x}) &\to& {\bar{\phi}}(t) + \delta \phi(t,{\bf x})\, ,\\
\chi(t,{\bf x}) &\to& {\bar{\chi}}(t) + \delta \chi(t,{\bf x})\,,
\end{split}
\eea
where $\bar{g}_{\mu\nu}$, $\bar{\phi}$, and $\bar{\chi}$ indicate the spatially averaged homogeneous background values. Since any matter field that might have existed before inflation is depleted by exponential expansion, one can assume the field $\chi$ starts from its vacuum state, { i.e.,} $\bar{\chi} = 0$ and $\delta \chi(t,{\bf x}) = \chi(t,{\bf x})$. From here onwards, $\delta\chi$ and $\chi$ will be used interchangeably. The homogeneous oscillating background $\bar{\phi}(t)$ satisfies the equation \eqref{eq:phi-evolution-background}. At linear order in perturbation theory, the equations of motion for $\delta\phi$ and $\chi$, following \eqref{eq:EoM-phi-chi1} and \eqref{eq:EoM-phi-chi2}, are
\bea\label{eq:delta-phi-EoM}
\delta \ddot{\phi} + 3H\delta\dot{\phi} + \bigg[ -\frac{{\boldsymbol\nabla}^2}{a^2} + \frac{\partial^2 V(\bar{\phi})}{\partial \bar{\phi}^2} \bigg] \delta\phi &=& 0\, ,\\
\label{eq:chi-EoM}
\ddot{\chi} + 3H\dot{\chi} + \bigg[ -\frac{{\boldsymbol\nabla}^2}{a^2} + g^2 \bar{\phi}^2 \bigg] \chi &=& 0\, .
\eea
Since the duration of preheating is small compared to the Hubble time $H^{-1}$, the growth of the perturbations are aptly captured by {\it Floquet analysis}, which is usually performed in a Minkowski background, which neglecting the Hubble friction \cite{Frolov:2010sz, Karouby:2011xs, Hertzberg:2014iza, Amin:2014eta}, {\it i.e.,} $a=1$ and $H=0$. In equations \eqref{eq:delta-phi-EoM} and \eqref{eq:chi-EoM} we substitute Fourier mode expansions of the fields
\be
\Phi(t, {\bf x}) = \int \frac{d^3 k}{(2\pi)^3} \Phi_k(t) e^{-i{\bf k}.{\bf x}}\, ,\
\ee
where $\Phi = \delta\phi, \chi$, and $\Phi_k=\delta\phi_k, \chi_k$ for a mode $k$ of fluctuations of the fields. These substitutions yield the following equations by setting $a=1$ and $H=0$.
\bea
\ddot{\delta\phi_k} + \bigg( k^2 + \frac{\partial^2 V(\bar{\phi})}{\partial \bar{\phi}^2} \bigg)\delta\phi_k &=& 0\, ,\\
\ddot{\chi_k} + \bigg( k^2 + g^2\bar{\phi}^2 \bigg)\chi_k &=& 0\, .
\eea
They constitute a set of two independent parametric oscillators, whose solutions are obtained numerically using Floquet theorem \cite{Floquet-1}
\bea\label{eq:floquet-theorem}
\begin{split}
\delta\phi_k &=& P^\phi_+(t) e^{\mu_k t} + P^\phi_-(t) e^{-\mu_k t}\, ,\\
\chi_k &=& P^\chi_+(t) e^{\nu_k t} + P^\chi_-(t) e^{-\nu_k t}\, ,
\end{split}
\eea
where $P^\phi_\pm(t)$ and $P^\chi_\pm(t)$ are quasi-periodic functions with periods matching the oscillation frequencies of the fields. The coefficients $\mu_k$ and $\nu_k$ called Floquet coefficients, determine the stability and instability regions of the oscillations. In general, the Floquet coefficients are complex.  For a given mode $k$, if the ${\rm Re}(\mu_k)>0, {\rm Re}(\nu_k)>0$ then the modes are in an unstable exponential growth phase, whereas for purely imaginary values of  $\mu_k, \nu_k$ the modes are in a phase of stable oscillations.

With an in-house developed code written in \texttt{Fortran} based on the algorithm presented in \cite{Amin:2014eta}, we have obtained the Floquet stability-instability regions shown in \ref{fig:floquet-mu}. The top and the bottom plots correspond to $\delta\phi_k$ and $\chi_k$, respectively. The plots are obtained for coupling constant $\xi^2=3.6\times 10^{-14}$. In each plots, the vertical axes are $\bar{\phi}/m_{\rm Pl}$, and the horizontal axes are $k/m$, where $m = \partial^2V(\bar{\phi})/\partial\bar{\phi}^2|_{\bar{\phi}_{\rm min}}$ is the mass of inflaton calculated at the potential minimum. The color gradient represents the value of the real part of Floquet coefficients ${\rm Re}(\mu_k)/m$ and ${\rm Re}(\nu_k)/m$. Brighter colors indicates regions where $\delta\phi_k$ and $\chi_k$ might undergo exponential growth. The figure corresponding to $\delta\phi_k$ displays broad self-resonance for modes $k<0.6m$. An external parametric-resonance corresponding to $\chi_k$ is also present for modes $k<0.2 m$. Although resonances are present for $\delta\phi_k$ and $\chi_k$, whether they are sufficiently strong to induce unstable growth in perturbations, especially due to the facts that the Floquet coefficients are of the order $\mathcal{O}(10^{-2}-10^{-1})$ and the width of the parametric resonance band is narrow, requires further investigation.

\begin{figure}[h!]
\center
\includegraphics[width=0.44\textwidth]{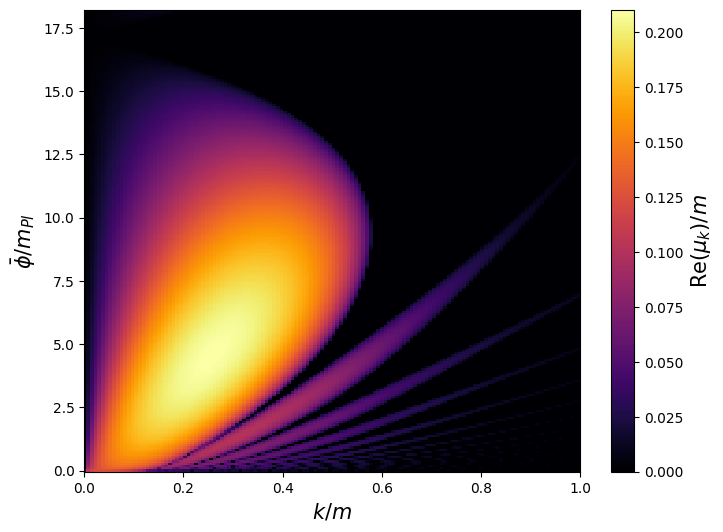}
\includegraphics[width=0.44\textwidth]{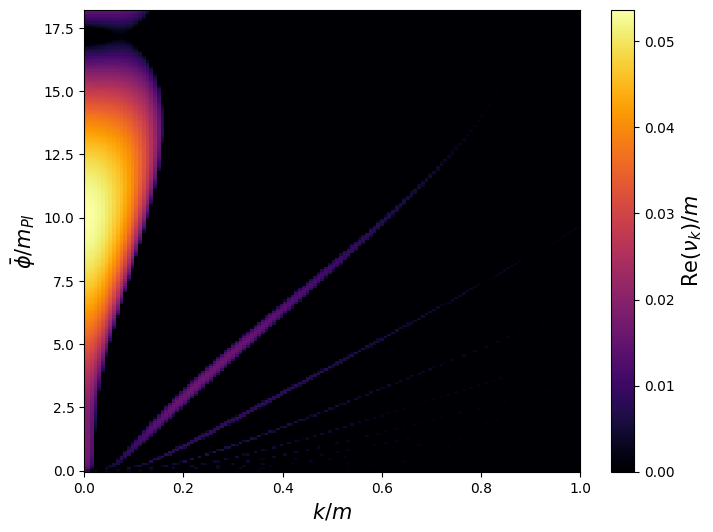}
\caption{The stability-instability regions from the Floquet analysis. The top panel shows the self-resonance of the inflaton and the bottom panel corresponds to the external parametric resonances of $\chi$. The brighter regions correspond to instability where $\delta\phi_k$ and $\chi_k$ experience exponential growth. The plots have been obtained for $\xi^2 = 3.6\times 10^{-14}$. }
\label{fig:floquet-mu}
\end{figure}

\section{Lattice Simulation \label{sec:lattice}}
The self- and external-resonances revealed in Floquet analysis encourage us to further explore the nonlinear dynamics of the fields through a lattice simulation. Among the several publicly available codes for the purpose \cite{Felder:2000hq, Sainio:2009hm, Frolov:2008hy, Easther:2010qz, Huang:2011gf}, we chose $\mathcal{C}$osmo$\mathcal{L}$attice \cite{Figueroa:2020rrl, Figueroa:2021yhd}. For a given set of $n$ dynamical fields $\{\phi \}$, 
\begin{align}\label{eq:HLattice-equations}
\begin{split}
& \ddot{\phi}_n + 3H\dot{\phi}_n - \frac{1}{a^2} {\boldsymbol\nabla}^2\phi_n + \frac{\partial V(\{\phi\})}{\partial \phi_n} = 0\, ,\\
& H^2 = \frac{1}{3m_{\rm Pl}^2} \bigg[ V(\{\phi\}) + \sum_n\bigg( \frac{1}{2}\dot{\phi}_n + \frac{1}{2a^2} |{\boldsymbol\nabla} \phi_n|^2  \bigg)  \bigg]\, ,
\end{split}
\end{align}
where $n$ (=2 in our case), and ${\boldsymbol\nabla}^2$ is the Laplacian operator, the $\mathcal{C}$osmo$\mathcal{L}$attice solves the following set of equations by discretizing them on a lattice.

The $\mathcal{C}$osmo$\mathcal{L}$attice implementation of any model starts with replacing the physical fields and the spacetime $(f, t, x^i)$, where $f=(\bar{\phi}, \bar{\chi})$, by dimensionless fields and spacetime $(\tilde{f}_\ast,\tilde{\eta},\tilde{x}^i)$ known as the {\it program variables}. The relations between the sets are
\be
\tilde{f} \equiv \frac{f}{f_\ast}\, ,\quad d\tilde{\eta} \equiv a^{-\alpha} \omega_\ast dt\, ,\quad d\tilde{x}^i \equiv \omega_\ast dx^i\, ,\nn
\ee
where $\alpha$ is a parameter, and $f_\ast, \omega_\ast$ are constants with dimension of energy.

Since $f_\ast$ and $\omega_\ast$ are rescaling factors, they are chosen so that the resulting numbers are close to unity. In our model the field value at which the inflation ends is $\bar{\phi}_{\rm end} = 15.87\mpl$ determined from the condition $\epsilon_V(\phi_{\rm end})=1$. Since the oscillatory phase occurs for field values larger than $\bar{\phi}_{\rm end}$, we set the initial amplitude of oscillation to $f_\ast=\bar{\phi}_{\rm end}$. To determine the initial frequency, the homogeneous potential is Taylor expanded about the minimum $\bar{\phi}=\bar{\phi}_{\rm min}$ as
\begin{align*}
V(\bar{\phi}) = V(\bar{\phi}_{\text{min}}) &+ \frac{\partial V}{\partial \bar{\phi}}|_{\bar{\phi}=\bar{\phi}_{\text{min}}}(\bar{\phi} - \bar{\phi}_{\text{min}})\\ &
+ \frac{1}{2}\frac{\partial^2 V}{\partial\bar{\phi}^2}|_{\bar{\phi}=\bar{\phi}_{\text{min}}}(\bar{\phi}-\bar{\phi}_{\text{min}})^2 + \cdots
\end{align*}
where the first two terms of the expansion vanish. Therefore, the equation of motion of the homogeneous background is
\be
\ddot{\bar{\phi}} +3H\dot{\bar{\phi}} + \bigg(\frac{\partial^2 V}{\partial\bar{\phi}^2}\bigg|_{\bar{\phi}_{\text{min}}} \bigg) (\bar{\phi}-\bar{\phi}_{\text{min}}) = 0
\ee
The equation suggests that value of the initial frequency is
\be
\omega_\ast = \sqrt{ \frac{\partial^2 V}{\partial\bar{\phi}^2}\bigg|_{\bar{\phi}_{\text{min}}} } \, .
\ee
which is equal to the mass of the inflaton $m$. Finally, we note that near the potential minimum, where the oscillation occurs, the dominant term of the potential is quadratic. Thus we chose $p=2$, which gives $\alpha=0$ \cite{Figueroa:2020rrl, Figueroa:2021yhd}.

In terms of the program variables, the potential and its first and second derivatives are given as
\begin{align}
\begin{split}
    \tilde{V}(\tilde{\phi})&=\frac{1}{f_{\ast}^2\omega_{\ast}^2}V(\tilde{\phi})\,\\
    &=\frac{1}{\bar{\phi}_{\text{min}}^2 m^2}\left( V_0 + \sum_{n=2}^{n=q} \frac{a_n}{n!} \bigg( \frac{\bar{\phi}_{\text{min}} \tilde{\phi}}{\phi_0} \bigg)^n \right)
\end{split}
\end{align}
\begin{align}
    \frac{\partial\tilde{V}}{\partial\tilde{\phi}} = \frac{1}{\bar{\phi}_{\text{min}}^2 m^2}\sum_{n=2}^{n=q} \frac{a_n}{(n-1)!} \bigg( \frac{\bar{\phi}_{\text{min}} }{\phi_0}  \bigg)^n \tilde{\phi}^{n-1}\\
    \frac{\partial^2\tilde{V}}{\partial\tilde{\phi}^2} = \frac{1}{\bar{\phi}_{\text{min}}^2 m^2}\sum_{n=2}^{n=q} \frac{a_n}{(n-2)!} \bigg( \frac{\bar{\phi}_{\text{min}} }{\phi_0}  \bigg)^n \tilde{\phi}^{n-2}
\end{align}
The interaction term between the fields is transformed as
\be
\frac{1}{2}\xi^2\bar{\phi}^2\chi^2 \to \frac{1}{2}\tilde{\xi}^2\tilde{\phi}^2\tilde{\chi}^2\, ,
\ee
where $\tilde{\xi}^2=\xi^2 \bar{\phi}_{\text{min}}^2 /m^2$.

We performed our simulations with an infrared cutoff $k/am=0.01$ and lattice size $N=128^3$. The results have been checked for two choices of coupling constant $\xi^2=3\times 10^{-10}$ and $\xi^2 = 3.6\times 10^{-14}$ and several arbitrarily chosen benchmark points shown in figure \ref{fig:planck-ns-r}. Different couplings and benchmark points yield qualitatively similar results. Here we present the results for the coupling constant $\xi^2 = 3.6\times 10^{-14}$ and benchmark shown in Table \ref{tab:benchmark}. In figure \ref{fig:phi-chi-osc} the evolution of the homogeneous inflaton background is shown. The background field $\bar{\phi}$ oscillates about the minima of the potential with a decreasing amplitude. The decrease in the amplitude may be attributed to the Hubble friction and dissipation of the inflaton's energy to the $\chi$ field. 

\begin{figure}[ht!]
\center
\includegraphics[width=0.45\textwidth]{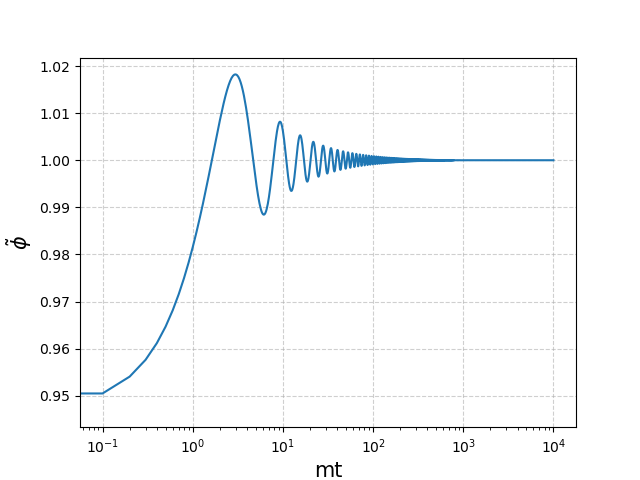}
\caption{Evolution of the background field $\tilde{\phi}$ for a coupling constant set to $\xi^2=3.6\times 10^{-14}$.}
\label{fig:phi-chi-osc}
\end{figure}

The scalar power spectrum of the fluctuations of the inflaton and the $\chi$ field is shown in figure \ref{fig:phi-power}. The color gradient represents the time of evolution, with darker shade indicating earlier times. In both figures, the power spectrum decreases as the universe expands. The amplitudes for small modes are not amplified by the resonances observed in the Floquet analysis. The absence of amplifications of smaller modes indicates that the occupation number of $\delta\phi$ and $\chi$ remain close to zero. The discrepancy between the lattice results and the Floquet analysis can be explained from the fact that, in the Floquet analysis, the expansion of the universe -- which suppresses the growth of fluctuations -- was neglected. Additionally, the magnitude of the Floquet coefficients is small and decreases near the minima of the potential, where the oscillation occurs. 
\begin{figure}[ht!]
\center
\includegraphics[width=0.47\textwidth]{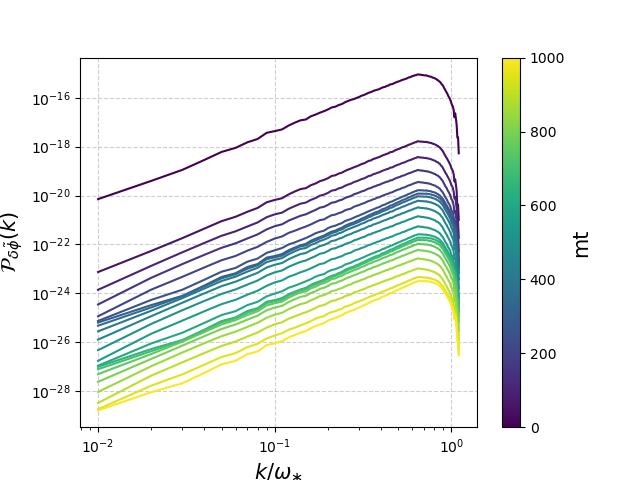}\\
\includegraphics[width=0.47\textwidth]{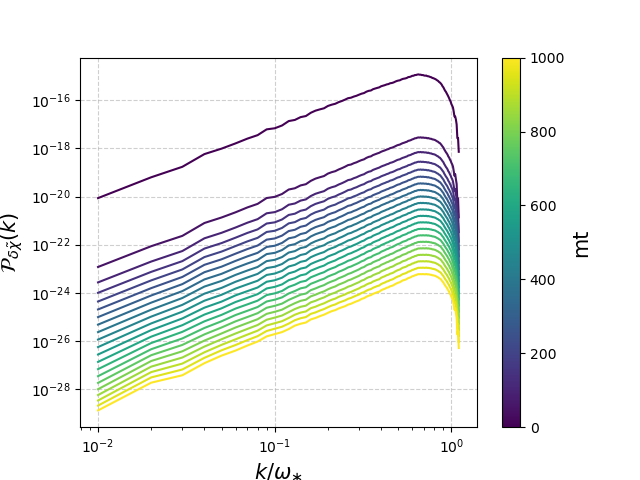}
\caption{Evolution of the power spectrum $\mathcal{P}_{\delta\tilde{\phi}}$ (above) and $\mathcal{P}_{\delta\tilde{\chi}}$ (below).}
\label{fig:phi-power}
\end{figure}

The initial inhomogeneities might still give rise to a stochastic gravitational wave background spectrum, which we now study. The gravitational waves are manifestations of the transverse-traceless (TT) and gauge-invariant tensor components of the perturbations, $h_{ij}$, on the background FRW spacetime \cite{Lifshitz:1945du, Bardeen:1980kt}
\be\label{eq:perurbed-FRW}
ds^2 = -dt^2 + a^2(\delta_{ij} + h_{ij})dx^i dx^j\, .
\ee
The $h_{ij}$ are function of space and time with $|h_{ij}|\ll 1$. The transverse-traceless (TT) conditions
\be
h^i_i = 0\, ,\quad \text{and}\quad \partial_i h_{ij} =0\, .\nn
\ee
leads to only two degrees of freedom for $h_{ij}$ and ensures that there is no interference between scalar and vector perturbations. From the perturbed metric \eqref{eq:perurbed-FRW}, the perturbed Einstein's equation follows $\delta G_{ij} = 8\pi G\delta T_{ij}$, where $G_{ij}$ is the Einstein tensor. It leads to the equation of motion for the tensor perturbations $h_{ij}$
\be\label{eq:EoM-hij}
\ddot{h}_{ij} + 3H\dot{h}_{ij} - \frac{1}{a^2} \nabla^2 h_{ij} = 16\pi G \Pi^{\rm}_{ij}\, .
\ee
where $\Pi_{ij}$ is the source of the gravitational waves satisfying the transverse-traceless conditions $\partial_i \Pi_{ij}=0$ and $\Pi_{ii}=0$. The source term is equal to the transverse-traceless part of the total anisotropic stress-tensor of both the inflaton and the $\chi$ field
\be
T_{\mu\nu} = \frac{1}{a^2} \bigg[ \partial_\mu \phi \partial_\nu \phi + \partial_\mu\chi \partial_\nu \chi + g_{\mu\nu} (\mathcal{L} - \langle p\rangle)  \bigg]\,,
\ee
where $\mathcal{L}$ is the Lagrangian in the action \eqref{eq:action}, and $\langle p\rangle$ is the homogeneous background pressure.

\begin{figure}[ht!]
\center
\includegraphics[width=0.45\textwidth]{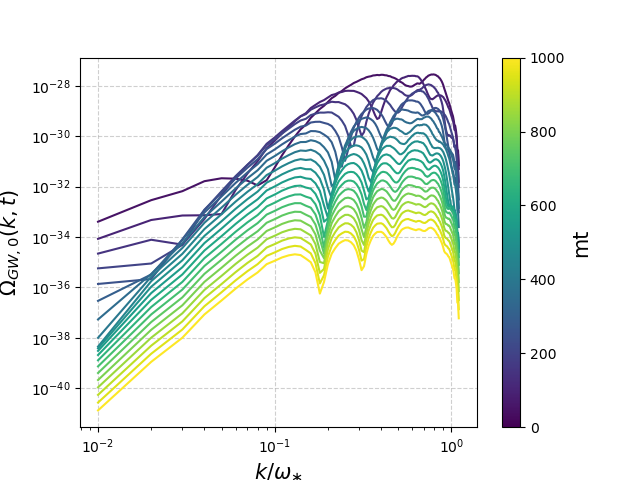}
\caption{The present-day gravitational waves background signal.}
\label{fig:gw}
\end{figure}

\begin{figure}[ht!]
\center
\includegraphics[width=0.39\textwidth]{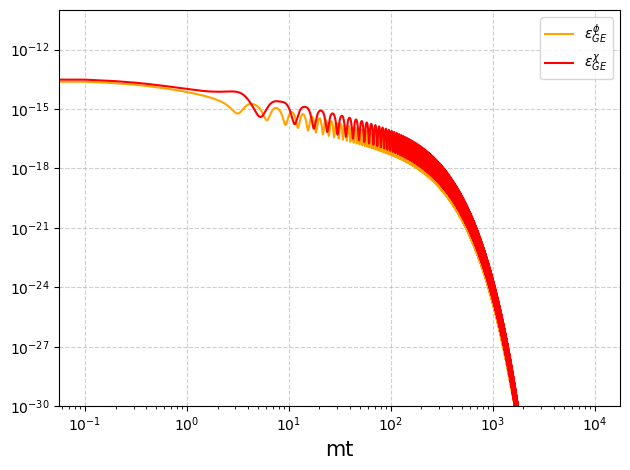}
\caption{The fractional gradient energy densities.}
\label{fig:gradient-energy}
\end{figure}

The solutions $h_{ij}(t,{\bf x})$ allow us to define the energy density of the gravitational waves
\be
\rho_{\rm GW}(t) = \frac{m_{\rm Pl}^2}{4} \langle \dot{h}_{ij}(t,{\bf x}) \dot{h}_{ij}(t,{\bf x}) \rangle\, ,
\ee
where $\langle \cdots \rangle$ indicates the average over a sufficiently large volume. The gravitational waves spectrum, normalized by the critical density $\rho_c$ of the universe, is defined as
\be
\Omega_{\rm GW} = \frac{1}{\rho_c} \frac{d\rho_{\rm GW}}{d\ln k}\, ,
\ee
where $f$ is the gravitational wave frequency, and $\rho_c$ is the critical density of the universe. The present-day gravitational wave background, $\Omega_{\rm GW,0} $ is shown in figure \ref{fig:gw}. As the figure shows, a stochastic gravitational wave is generated in this model likely due to the initial inhomogeneities. This background is below the present limit set by CMB+BAO+Lensing+$^2$H observations \cite{Pagano:2015hma}. However there is no peak in the spectrum due to resonance instabilities during preheating. It means that the initial inhomogeneities do not grow. Inhomogeneities are associated to the gradient term of the action.  In terms of the program fields $\tilde{f}$ it is written as
\be
\widetilde{G}_{\tilde{f}} = \frac{1}{2a^2(\tilde{t})} \sum_i \bigg(\frac{\partial \tilde{f}}{\partial\tilde{x}_i} \bigg)^2\, ,
\ee
from which the fractional gradient energy density is define as
\be
\varepsilon_{\rm GE}^{\tilde{f}} = \frac{\widetilde{G}_{\tilde{f}}}{\tilde{\rho}}\, ,
\ee
where $\tilde{\rho}$ is the total energy density of the two fields combined. In figure \ref{fig:gradient-energy}, the result for the gradient energy density for the two fields are shown. At the start of the simulation, inhomogeneities are generated, but it decreases rapidly. So significant nonlinearities do not develop in this model.

\section{Summary \label{sec:summary}}
In the standard model of cosmology, the concept of inflation elegantly explains the flatness problem and the horizon problem, and predicts a primordial density perturbation that is largely consistent with observations in the cosmic microwave background. Among the many single-field inflationary potentials, the large-field regularized hilltop model remains consistent with the latest Planck+BAO+BICEP/Keck data. In this paper, we explore preheating and gravitational waves in this model with an additional scalar field coupled to the inflaton. We perform a Floquet analysis using linear perturbation theory. The analysis shows that the small modes of inflaton's fluctuations, $\delta\phi_k$, at the end of inflation experience self-resonance, while external parametric-resonance occurs in $\delta\chi_k$. However, we find that the resonances are not strong enough to drive exponential growth of the fluctuations; hence, the occupation number of $\delta\phi_k$ and $\delta\chi_k$ remain close to zero. This is corroborated by results from a lattice simulation, which shows that in the power spectrum of the scalar fields, small modes do not experience any amplification. However, the lattice simulation shows that a stochastic gravitational wave background, arising from initial inhomogeneities, is present but remains below the current observational limit.

\section*{Acknowledgment}
DD would like to thank Mustafa Amin and Ramkishor Sharma for helpful communications, Chandan Hati for kind hospitality during a visit to IFIC Valencia, and Hans-Peter Buechler for support during a stay at the University of Stuttgart. DD also acknowledges support under the SERB SRG grant (Sanction Order No. SRG/2023/001318) and the IIIT Hyderabad Seed Fund (No. IIIT/R\&D Office/Seed-Grant/2021-22/013).
 
\end{document}